\documentclass[11pt, a4paper]{article}

\usepackage[utf8]{inputenc}
\usepackage[margin=1in]{geometry}
\usepackage{authblk}
\usepackage{hyperref}
\usepackage{amsmath}
\usepackage{amsfonts}

\title{Astrophysics in the Era of Artificial Intelligence powered by  Large Language Models}

\author{A. R. Rao}
\affil{Department of Astronomy and Astrophysics, \\ Tata Institute of Fundamental Research, Mumbai 400005, India}
\affil{Email: \href{mailto:a.raghu.rao@gmail.com}{a.raghu.rao@gmail.com}}

\date{\today}

\begin{document}

\maketitle

\begin{abstract}
In recent times there is considerable unease about the way  Artificial Intelligence (AI) powered by large language models is used to do astrophysics, and, assuming that these AIs will get far better in future,  several alarming situations are also discussed. 
In a  ``white paper,'' Hogg argues that extreme cases like allowing AI to write all astrophysics papers or strictly controlling AI are not good choices and posits that it is difficult to adopt moderate policies. I examine the impact of AI on astrophysics by making two rather extreme assumptions:  first, that the modern astrophysicist has become increasingly preoccupied with the administrative task of publication over the act of physical discovery (astrophysicists are busy writing papers and have stopped doing astrophysics); and second, that while AI possesses the capacity to synthesize literature and generate manuscripts, they lack the intrinsic ability to perform original astrophysical inquiry (AI can write papers but cannot do astrophysics).  I provide some plausible reasons for adopting these assumptions, but, more importantly, extrapolate the implications of these assumptions to argue that AI  can have a very positive impact on astrophysics. By situating AI as agents of publication rather than discovery, the focus of the human researcher can be directed towards fundamental astrophysics, leading to very positive growth in the field.
\end{abstract}

\section{Introduction}

Ever since the dramatic entry of Artificial Intelligence powered by large language models (for brevity,  referred collectively as AI) into the public conscience by ChatGPT, they are increasingly used in every walk of life where information and organising information (using codes) are used. Astrophysics is no exception. Doing astrophysics involves collecting information from published literature, using increasingly sophisticated and, most often, publicly available data, using clever codes to understand these data, and associating the results to the current knowledge in the field and arriving at meaningful conclusions. It is obvious that AI would be very useful for many of these tasks, and it is not inconceivable to imagine that in the near future a single line instruction like ``examine all publicly available data from this particular observatory and identify a new changing look AGN and write a ApJ style paper'' would result in a near-ready paper not inferior in quality to any published work in this particular area. 

This has led to some alarm. In a paper titled ``Why do we do astrophysics?"  David W. Hogg  \cite{Hogg2026} addresses the profound shift in the field of astrophysics caused by the rapid advancement of AI and the increasing professionalization of astronomical data production. Hogg notes that AI is beginning to demonstrate the capability to design, execute, and even referee scientific projects, particularly on the data-science side of the discipline. This technological leap coincides with a trend where large-scale missions, such as the ESA's Gaia Mission or NASA's JWST, are increasingly built and operated by engineering professionals and defence contractors rather than research-active astronomers. Hogg posits that the extreme approach of fully embracing machines to perform the bulk of astrophysical research would lead to the ``death" of astrophysics as a human activity because it is practically impossible for humans to coexist with AI that could eventually produce papers at millions of times the human rate. On the other hand, the extreme case of completely banning the use of AI in astrophysics is also bound to fail. 

I argue here that many of the alarming scenarios that are talked about regarding the impact of AI in doing research come by equating doing astrophysics to writing papers.  Here I want to make a clear distinction between the act of ``writing science" and the act of ``doing science". I further argue that, in recent times, particularly after the explosive growth of the internet, which enabled the easy availability of astrophysical data,  software, and literature, the major preoccupation of a professional astrophysicist has been in the various acts of professionalising the act of ``writing science", thus leading to a gradual decline of  the act of ``doing science". Hence, the likely disruption that would be caused by AI doing the bulk of  the act of ``writing science" will automatically force astrophysicists to specialise in the act of ``doing science".
 To drive home my point, I make a couple of simplistic extreme assumptions. Examining the implications of these assumptions leads to the conclusion that the impact of LLMs on doing astrophysics is an extreme disruption, and very beneficial for the overall well being of humanity. 

\section{The Two Assumptions}

\subsection{Assumption 1: Astrophysicists write papers but don’t do astrophysics }

I am aware that this is an extreme statement, perhaps a bit insulting to a vast number of sincere and serious astrophysicists doing a meticulous job of recording the fruits of their labour as quality research papers. Rather, this is an attempt to reinforce and assert what doing astrophysics means: like any branch of science, it is all about acquiring new knowledge with a special emphasis on creative insights. Creativity is the key aspect of this process, and it becomes people centric in the sense that the practice of a creative activity increases the creative talent of the practitioner, and hence it is an ever increasing positive feedback loop of increasing creativity.

The product of the creative activity of astrophysics is most often recorded as a research paper. But, in recent times, there is a heavy reliance on the number of papers and their citations to measure not only the quality of a researcher but also the quality of research institutes and research facilities. Further, writing a research paper is heavily used as an educational tool (quite often with a prescribed number of research papers to qualify for a PhD degree). The overemphasis on product (research papers) rather than the process has, in my opinion, resulted in a steady decrease in the amount of ``astrophysics’’ recorded in a typical research paper and has crossed a hypothetical lower limit to enable me to assert that the majority of research papers do not contain a significant amount of astrophysics and hence astrophysicists mainly write papers rather than doing astrophysics. Further, the wide use of the internet in this century has also resulted in an explosive growth in the number of papers, thus demanding the attention of a typical astrophysicist to a multitude of related tasks (refereeing, editing, etc.) which do not necessarily enhance the creative qualities of a researcher.

There is anecdotal evidence for this phenomenon. Antonucci\cite{Antonucci2013}, while reviewing our understanding of Quasars, points out that ‘Quasars still defy explanation’ and ‘...having given up on understanding AGNs, the community now focuses on the more modest goal of counting them’. It is also noted that in recent times, there is ‘...a lack of progress in the long term and deep thoughts’ in astronomy and astrophysics \cite{Rao2015}. In the area of accretion onto black holes, ‘having given up on understanding accretion onto black holes, the community now focuses on the more modest goal of categorising QPOs’  \cite{Rao2024}. While discussing the effect of public science on corporate R\&D, Arora et al. \cite{Arora2023} conclude that, as far as the effect on industry goes, the university-led R\&D effort did not significantly impact industrial productivity.

The evidence is anecdotal and subjective, and hence, for the purpose of the present discussion, I have stated this merely as an assumption to explore the possible impact of AI on doing astrophysics. On a formal note, one can reframe the assumption as follows. The number distribution of research papers in astrophysics as a function of the astrophysical quality q of the paper can be assumed to be a power law starting from q$_0$ and index $\alpha$. In recent times, the index has become steeper due to the large number of quick papers. The first assumption asserts that the average value of q these days is very low, and hence the majority of the papers lack significant astrophysical content.

\subsection{Assumption 2: AI can write papers but cannot do astrophysics.}

The rapid rise of AI and its sophistication is primarily the result of the never ending Moore’s Law, by which the number of transistors on a microchip doubles approximately every two years, thus giving an accelerated performance of all digital tools. With the vast computational power that is available, most of the routine tasks can indeed be digitised and accomplished in much shorter time than what humans can do. The creativity aspects of humans, developed over millions of years of evolution, perhaps could be difficult to emulate in a digital landscape. 

Again, the reasons are heuristic and lack rigorous proof and, hence, I merely use it as an assumption that AI can write papers but cannot do astrophysics to investigate the impact of AI on astrophysics. In the formal context discussed earlier, we can say that  AI can write papers up to a critical quality q$_c$. Though q$_c$ can be an increasing function of time (AI getting better with time), it is assumed that there exists a quality factor q$_h$ beyond which only humans can write astrophysical papers.

\section{Impact of AI on astrophysics}

Armed with these two assumptions of extreme nature and along with a dose of optimism, now I proceed to speculate on the future of astrophysics in the era of AI. It is implicitly accepted, as a corollary to the two assumptions, that astrophysics papers and other related metrics like citations will completely loose the authenticity to use them as a measure of the process called astrophysics. 

\subsection{Impact on established researchers}

The overwhelming stress on the importance of number of papers and their citations in the mental make up of an established researcher in astrophysics has generated an extremely negative vicious cycle of a vast amount of related, but not necessarily creative, efforts, like competitively writing proposals for grants and time in observatories, hiring students and postdocs (and training them), having multitudes of collaborations which maximise the number of papers, and a host of other research paper related activities like correcting manuscripts, revising, editing, and other similar activities.

If the above phenomenon is one of the leading causes for our first assumption that astrophysicists write papers but do not do astrophysics, the fact that AI is able to independently write research papers of qualities similar to those produced now can result  in an extreme disruption in the process of writing papers: research papers can no longer be used as a metric to measure the quality of an astrophysicist.  Let us assume, as a first approximation, that astrophysicists would be forced to concentrate more on the astrophysical content of research papers. Recall our second assumption that AI can write papers but cannot do astrophysics: a substantial increase in the content of astrophysics in research papers will make them stand out among the AI-written research papers. 

Postponing a discussion on what could possibly replace research papers as metrics of the astrophysical abilities of a researcher to a later section, let us first focus on the impact on an astrophysicist in the era of AI. First of all, most of the routine stuff that a researcher generally gets done through a graduate student and a post-doc can be easily done by AI. Hence, the researcher can fully concentrate on the core astrophysical ideas. The rush to get the prime data from the latest observatory to identify some peculiar object in the sky will abate because, as routine tasks of building instruments and writing pipe-line codes are already done by professional contractors, reporting of novel events in the sky can be done by AI enabled automatic tools. Research in astrophysics, instead of being a zero sum game (there are only a limited number of new phenomena to be reported from an observatory and the first person to get there gets the credit), can truly metamorphose into a  creative endeavour dedicated to solving many mysteries out there in the universe. A set of professional astronomers, enabled by AI to take care of the routine tasks, can indeed transform astrophysical research to an unprecedented level.

Of course, it might take quite a while for the disruption to fully materials, but, in the current literature too, there are indications of some positive movement. 

De Rujula, a theoretical physicist advocating the Cannonball model for Gamma-ray Bursts (GRBs), recently \cite{Derujula2026} posted an article in the archives: ``Human versus Artificial Intelligence; various significant examples in astrophysics’’ which is essentially a discussion on GRB Models. What is interesting is that by using Perplexity.ai, De Rujula could get a complete comparison of the models and, as he points out, ``except for this abstract, two footnotes and two other references to standard  and CB-model articles and talks, all of what follows is, verbatim, what the cited AI “opines" ". 

In another recent article Dr. Nicholas E. White, perhaps best known for his efforts as founding director of the High Energy Astrophysics Science Archive Research Center (HEASARC), proposes a new ``Hybrid" Roche-Lobe Overflow (RLOF) model to resolve the long-standing geometric and kinematic mysteries of the Galactic X-ray binary Cygnus X-3 and concludes that Cygnus X-3 is a local, stable analog to Ultraluminous X-ray Sources \cite{White2026}. Generally, such papers are written with multiple authors (students and post-docs) who step in to do laborious calculations and/or coding. What is remarkable about this paper is that it is a single author paper and the author charmingly acknowledges the use of Gemini (Google) and ChatGPT for assistance with editing and formatting the manuscript. This example does illustrate the point that serious astrophysics can be done using AI without the help of working hands provided by students and post-docs.

In summary, serious astrophysicists, who were earlier tied down to a vast amount of  routine tasks of enabling doing astrophysics, can, in principle, fully concentrate on doing astrophysics. Of course, it might take some time for the deeply ingrained habits of going after the number of publications to change, but the fascination of doing astrophysics is something firmly embedded in a large number of practising astrophysicists, and the disruption caused by AI can push such serious astrophysicists into a more meaningful routine.

\subsection{Impact on post-docs}

Post-doc work, as a profession, most likely will cease to exist.

Much has been written about the exploitative nature of the post-doc work in astrophysics, where the number of post-docs far exceeds the number of academic posts available in astrophysics. The nature of post-doc work eerily resembles the survival of the Chicago gang``Black Disciples"  described in `Freakonomics’ by Stephen J. Dubner and Steven Levitt \cite{Freakonomics}. These authors argue that the system relied on a ``strictly hierarchical, corporate-style business model’’ that incentivized low-level workers to accept extreme risks for almost no pay. Despite low wages,``foot soldiers" remained in the gang because they viewed their roles as ``high-stakes apprenticeships’', fuelled by the hope of ascending to the lucrative next step where leaders lived in luxury. This tournament-style economic structure allowed the organization to maintain a massive labor force that bore the brunt of the work, effectively subsidizing the wealth of the top tier while ensuring the gang's operational longevity through a constant supply of desperate, aspiring talent.

Well, in astrophysics, the ``supply of desperate, aspiring talent’’ is fuelled by the romantic picture of doing astrophysics and also, astrophysics being an international effort, finding the talents from far corners of the world, quite often from poorer sections of society (youngsters from a rich society will hesitate to take up such a high-risk career). This style, inevitably, can also result in reducing the astrophysical quality of the research output, and also in reducing the astrophysical training of an average post-doc. That is, a typical post-doc worker does most of the routine jobs and hence gets trained in the specific jobs of the domain, thus reducing the job prospects in other related areas. 

If we assume that most of the routine jobs can be done by AI, there is no incentive for a researcher to hire a post-doc, and there is no incentive for a post-doc to do routine jobs that can be easily done by AI. 
The `grinding’ typically done by a post-doc to acquire the finer nuances of doing astrophysics most likely will have to be done on the job. The typical post-doc stage of a research career of an astrophysicist, most likely, will be obtained during the tenure track of a research career (thus removing the imbalance between the number of post-docs and the number of academic jobs - a tenure-tracked person will invariably remain in the field, perhaps moving into a lower strata of an academic institute, if found unsuitable).

\subsection{Impact on PhD students}

The fate of PhD students most likely will follow that of the post-docs.

The ``Chicago gangster syndrome’’ is much more acute in the case of PhD students. It is not uncommon to hear an established researcher say that he has a very good idea, but is only waiting for a good graduate student to execute it. The reliance on PhD students and post-docs to carry out routine tasks has resulted in an eco-system where an academician’s primary job is to distribute work and monitor the smooth flow of the publication machinery. The end effect of this eco system is an ever-decreasing quality of the papers and an ever-decreasing attempt by astrophysicists to do serious astrophysics, being content to churn out a large number of papers. This system could sustain because the measure to judge the quality of an astrophysicist is mainly based on the number of publications (quality of these papers affecting only marginally, so long as the papers cross certain basic criterion like the journal in which it is published). 

The ability of the AI to turn out papers up to a certain quality will cause a huge disruption to this eco system. The number of papers cannot be a measure of the ability of an astrophysicist. There is no incentive for an astrophysicist to invest time with novice PhD students. Till recently a PhD in astrophysics could also find employment in data science related areas due to the heavy reliance of astrophysics in data analysis techniques: AI is causing a huge disruption in data analytics and a PhD in astrophysics with only the domain knowledge of rudimentary data analysis will find it very difficult to find alternate employment and hence there will be less incentive for an aspiring student to do a routine PhD in astrophysics.

The net result of the disruption caused by the ability of AI to turn out routine astrophysics papers most likely would be forcing astrophysicists to concentrate more on core astrophysics and for students to join the astrophysics stream only with a reasonable assurance of getting a job in astrophysics. One possible outcome of this disruption is that an astrophysics PhD would be obtained on the go in the job as an astrophysicist. 

\subsection{Impact on  research at the masters level}

Writing research papers has a good educational value. It worked both ways: the student gets an independent assessment of their abilities and an avenue to increase the prospects of their future career (mostly to get good PhD positions), and the faculty members amplify their number of publications. With most routine papers being written by AI, there is less incentive for the faculty to get routine papers and even less incentive for the students to go for a PhD. The rush for students to get research papers will diminish, and the evaluation of the master's student will rely more on traditional methods like marks in exams and evaluation by the teachers.

\section{Evaluation of the capabilities of astrophysicists}

If research papers and their citations lose their authenticity to measure the productivity of an astrophysicist due to the possible deluge of AI written research papers, some other mode of evaluation need to evolve. The over dependence on independently measurable matrices like numbers and citations essentially stems from the belief that such objective measures remove bias in other forms of subjective evaluation. But it is quite evident that these so-called objective measures can also be inflated or gamed, and hence some form of quasi-subjective measure of the quality of an astrophysicist, which is robust and appears to be fair, can emerge in the future.

I sketch below one possible alternative. Although it is highly speculative, a variant of this method could emerge in the future. 

Peer review is a time tested method in evaluating the quality of a research paper. An open peer review system, instead of the currently prevalent anonymous peer review system (which can be distorted) can indeed form the benchmark for the evaluation of the quality of an astrophysicist. Such a collective subjective system, which can be deemed to be objective to the level of the average objectivity of the majority of the participants, can be robust, particularly since the AI disruption will make astrophysicists concentrate more on astrophysics and perhaps sharpen their objectivity. Other alternate systems like social media likes can be swayed by passing trends. It is indeed quite satisfying for a practising astrophysicist to be openly peer reviewed by contemporaries. One can speculate here that all papers are automatically rated by AI and endorsed by an open peer review system where the readers opine on the content and the quality of the opinion and the quality of the opinion maker are automatically recorded for all to see so that a collective robust system of measuring the quality of an astrophysicist emerges (perhaps like the rating system for the chess players). 

Then, why would astrophysicists write papers when AI can write papers of a similar quality?  An astrophysicist records his current astrophysical thinking and its conclusions as a research paper, assisted by AI and mostly as a single-authored publication. Rating of this paper by an AI-assisted peer review system (say ranked 1 to 10, 10 being the maximum) would be immediately available. The purpose of writing papers is to record the astrophysical training, and once an astrophysicist reaches a quality factor n, the effort will be to go to the next level. At any given time, if AI can write papers up to a quality of, say, 5, all papers with quality  below this number would be mostly for self training and all efforts by astrophysicists would be to get to higher-quality papers. 

\section{Funding for astrophysics and jobs for astrophysicists}

Funding for astrophysics came primarily from three sources. 

The national pride of competing in space provided a vast amount of resources in space sciences, and astrophysical research coupled to space astronomy hugely benefited from this. The second source of funding is the natural fascination to space and the unknown in the general population, resulting in a positive feedback to the funders for eye-catching observatories and astrophysical research. The third source of funding is from the education sector, funded jointly by students, the state and also by philanthropists. 

The drawbacks in the first two funding avenues are twofold. Firstly, the projects need to be tailored to the availability of the resources rather than the requirements of the astrophysical pursuits. For example, if a country wants to flaunt her space achievements by building a space station, the proposals will be necessarily cater to the available resources rather than astrophysical requirements. Secondly, the fund givers want a measurable return for their investment, and they relied heavily on the metrics of publications, the number of papers and the number of citations. Quite often, researchers get paid to get publications from a given observatory. This could also be one of the reasons for the heavy reliance on publications to evaluate an astrophysical research. 

Both of the above-described drawbacks inevitably have the potential to dumb down the astrophysical content of a publication. If AI disrupts the industry of publications, astrophysicists need to find a way to be more professional in the way they project themselves and can tap this funding more meaningfully. 

The funding potential for astrophysicists in the education industry is quite large and would be sustainable even in the era of AI. Right now, a typical astrophysicist is valued in the education sector mostly by their ability to attract funds. Apart from this, there is value in the core method of doing astrophysics in the field of education. The ability to think deeply, ponder over the unknown, and come to some fascinating conclusions is generally uplifting to a human mind and sooner or later astrophysicists will be forced (once the disruption to the paper writing industry by AI is complete) to appropriately monetise their astrophysical abilities in the education sector. Some of the core educational requirements (like cohesive articulation, mathematical abilities, management abilities, knowledge of core subjects like Physics, computational abilities) are also the ingredients of a good astrophysical knowledge base, and it is quite feasible for a astrophysicist to impart this knowledge without seriously affecting the core astrophysical abilities. 

Of course, such methods are currently  prevalent in the education sector even now, but the effectiveness is quite compromised by shifting the focus of the astrophysicist from doing astrophysics to getting publication (and funding). Hopefully, in the future, due to the decreased emphasis on publication and evolving a reasonable metric for evaluating the quality of an astrophysicist, a new equilibrium could be reached, primarily based on the quality of the astrophysics and its direct utility to empower the human mind.

Can AI completely disrupt the education sector too and make the job of astrophysicists in the education sector redundant? Very, very unlikely. As has been found in primary education, there is no substitute for human touch and technology in education can only enhance the efficacy of the educator rather than replacing. AI can disrupt and make redundant many professions, but  many human activities like caregiving, educating, dreaming big and pondering over the unknown will surely remain the sole realm of humans, vastly assisted by AI.

\section{Conclusions}

The rise of AI does not signify the end of the astrophysicist but rather an opportunity for a professional renaissance. By offloading the increasingly heavy burden of manuscript production to automated systems, we can redirect human curiosity toward the unsolved mysteries of the cosmos. The ability to perform astrophysics is a skill to be nurtured, and AI may provide the breathing room necessary for the next generation of scientists to focus on depth over volume.

While X-ray spectral fitting using the XSPEC package by $\chi^2$ minimisation, quite often the program stops at some local minimum, and it requires some nudge by changing some parameters drastically for the program to reach a better and reasonable value of minimum $\chi^2$. In human endeavour too the multiple economic, social and other pressures can result in a local equilibrium. Some disruptions can move the system to a better equilibrium.

The easy availability of funds and the refinements of the methods of writing papers has resulted in recent times a deluge of astrophysics research papers, as argued here, primarily decreasing the quality of astrophysics. The ability of AI to write astrophysics papers of certain quality will disrupt this equilibrium. In this article it is romantically imagined that a new equilibrium with vastly improved quality of astrophysics will emerge in the future. 

Peiris \cite{Peiris2026} also has pointed out the futility of the current incentive structures: the ``publish or perish" culture prioritising volume over quality, leading to ``thin incremental work" and paper mills and advocates a shift in hiring and promotion away from volume-based metrics toward genuine intellectual contributions. In this article, however, I have emphasised the catastrophic effects of non-academic activities on the mental make-up of an astrophysicist. The cumulative effort required to sustain “thin incremental work” may have stunted the collective growth of astrophysics in the community. It is emphasised here that a continuous nurturing of the astrophysical qualities throughout one's career is highly essential, and the disruption caused by AI can be used towards this goal.
 
Finally, why did I write this article ?

Well, I really enjoyed the process of collecting my thoughts. Putting it all together indeed clarified many of my ideas too.

\section*{Acknowledgements}

I thank Claude.ai and Google Gemini for help in formatting the paper, collecting some pertinent information and other similar assistance. I, however, have gone through every sentence in this article and, to the best of my ability, vouch for their correctness.



\end{document}